\documentclass[letterpaper, 10 pt, conference]{ieeeconf}  

\IEEEoverridecommandlockouts                              

\usepackage{graphicx}
\usepackage{amsmath}
\usepackage{xcolor}
\usepackage{ragged2e}
\usepackage{amssymb}
\usepackage{tabularx}
\usepackage{booktabs}
\usepackage{subfig}
\usepackage{empheq}
\usepackage{stackrel}
\usepackage[font={small,it}]{caption}
\usepackage{hyperref}
\usepackage{url}

\newcommand{\leftrarrows}{\mathrel{\raise.75ex\hbox{\oalign{%
  $\scriptstyle\leftarrow$\cr
  \vrule width0pt height.5ex$\hfil\scriptstyle\relbar$\cr}}}}
\newcommand{\lrightarrows}{\mathrel{\raise.75ex\hbox{\oalign{%
  $\scriptstyle\relbar$\hfil\cr
  $\scriptstyle\vrule width0pt height.5ex\smash\rightarrow$\cr}}}}
\newcommand{\Rrelbar}{\mathrel{\raise.75ex\hbox{\oalign{%
  $\scriptstyle\relbar$\cr
  \vrule width0pt height.5ex$\scriptstyle\relbar$}}}}
\newcommand{\longleftrightarrows}{\leftrarrows\joinrel\Rrelbar\joinrel\lrightarrows}

\newbox\tempbox

\newbox\tempbox

\title{\LARGE \bf Core-shell enhanced single particle model for LiFePO$_\mathbf{4}$ batteries}

\author{Aki Takahashi$^1$,  Gabriele Pozzato$^1$,  Anirudh Allam$^1$,  Vahid Azimi$^1$, \\Xueyan Li$^{2}$, Donghoon Lee$^{3}$,  Johan Ko$^{3}$, and  Simona Onori$^{1,*}$
\thanks{$^{1}$ Energy Resources Engineering, Stanford University, Stanford, CA.}
\thanks{$^{2}$ LG Energy Solution Michigan, Troy,  MI.}
\thanks{$^{3}$  LG Energy Solution, South Korea.}
\thanks{$^{*}$ corresponding author  {\tt\small sonori@stanford.edu}}}

\begin{document}
\maketitle
\thispagestyle{empty}
\pagestyle{empty}
\begin{abstract}
In this paper, a novel electrochemical model for $\mathbf{LiFePO_4}$  battery cells that  accounts for the positive particle lithium intercalation and deintercalation dynamics is proposed.  Starting from the enhanced single particle model, mass transport and balance equations along with suitable boundary conditions are introduced to model the phase transformation phenomena during lithiation and delithiation in the positive electrode material.
The  lithium-poor and lithium-rich phases are modeled using the core-shell principle, where a  core composition is encapsulated with a shell composition.
The coupled partial differential equations describing the phase transformation are  discretized using the finite difference method, from which  a system of ordinary differential equations written in state-space representation is obtained. Finally, model parameter identification is performed using experimental data from a 49Ah LFP pouch cell. 
\end{abstract}

\section{Introduction}
Recently, $\mathrm{LiFePO_4}$ (LFP) batteries have received significant attention as the battery chemistry to use in electric vehicles (EVs) and other energy storage applications for their long-lasting,  safe, and environmentally-friendly and less costly materials.
Two phases exist during lithiation and delithiation of $\mathrm{LiFePO}_4$ electrodes: one Li-rich and one Li-poor \cite{Yamada_2005,delmas2011lithium}.  As a result of the coexistence of the two phases,  the positive electrode open circuit potential (OCP) shows a plateau and additionally,  the different composition of the two phases results in voltage hysteresis. 
A model that incorporates the description of the two phases can help understand lithium intercalation and deintercalation in $\mathrm{LiFePO}_4$ electrodes and, ultimately,  gain information on the electrochemical states.   Physics-based models proposed for this chemistry are a few.  In the many-particle model,  lithium is exchanged between individual particles and sequential lithiation and delithiation is demonstrated \cite{dreyer2010thermodynamic}.  However,  kinetic and transport equations are ignored,  making this model unsuitable for high C-rate or when a careful description of the electrochemical states is needed.  
In \cite{srinivasan2004discharge},  the core-shell approach is proposed to model phase transitions in LFP batteries.  At the cost of adding some negligible complexity to the electrochemical model\,--\,namely,  a mass balance equation\,--\,this technique allows for a detailed description of the battery intercalation and deintercalation dynamics.  This approach assumes isotropic diffusion in the positive particles,  while using a shell and core phase interacting via a moving boundary to describe phase transitions.  To account for the path dependence of battery operation,  the core-shell model prescribes different phases to the core and shell depending on whether the battery is charging or discharging.  Successful implementation of this core-shell model has been demonstrated in \cite{li2015modeling,koga2017state} for the pseudo-two-dimensional (P2D) model and single particle model (SPM),  respectively. In this paper, a core-shell enhanced single particle model (ESPM) is formulated.  Two phases, one poor and one rich in lithium, are used to describe the change of phase in the positive electrode, where the core of the cell is assumed to be at one phase and covered by the surrounding shell phase upon intercalation/deintercalation.  The proposed modeling strategy allows also to model the one-phase at the beginning (and at the end) of charge (and discharge).  A careful discretization of the electrochemical partial differential equations (PDEs) is proposed and used to convert the model into a system of ordinary differential equations (ODEs),  allowing for an effective numerical solution.  Moreover,  an optimization problem is formulated to identify the unknown model parameters.  
The remainder of the paper is organized as follows. Section \ref{sec:governingeq} describes the governing equations for  intercalation and deintercalation in LFP batteries and formulates the core-shell ESPM model.  In Section \ref{sec:numericalsol},  core-shell model equations are discretized and converted into a system of ODEs.  Section \ref{sec:paramid} describes the parameter identification procedure.  Finally,  in Section \ref{sec:results}  the model performance over charge and discharge experimental data is shown.

\section{Cell governing equations}\label{sec:governingeq}
\subsection{Physical principles}\label{sec:physics}
When a lithium-ion battery is being discharged ($I>0$), positively charged lithium ions ($\mathrm{Li}^+$) move from the negative to the positive electrode, whereas the lithium ions move back from the positive electrode to the negative electrode in charging ($I<0$). 
Deintercalation (intercalation) occurs when lithium ions leave (enter) one electrode.  
In LFP batteries,  the intercalation and deintercalation process in the LiFePO$_\mathbf{4}$ electrode is described by the following chemical reaction: $\mathrm{FePO_4 + Li^+ + e^-  \stackrel[discharge]{charge}{\longleftrightarrows} LiFePO_4}$. During discharge ($\rightarrow$),  lithium intercalates into the positive electrode and $\mathrm{LiFePO_4}$ is formed.  During charge ($\leftarrow$),  an oxidation reaction takes place and $\mathrm{FePO_4}$,  $\mathrm{Li}^+$,  and $\mathrm{e}^-$ are formed.  According to \cite{dreyer2010thermodynamic},  in LFP batteries the positive particle experiences the formation of two phases: a Li-poor phase, denoted $\alpha$, and a Li-rich phase, denoted $\beta$. 
\begin{figure}[!t]
\centering
\includegraphics[width = 1\columnwidth]{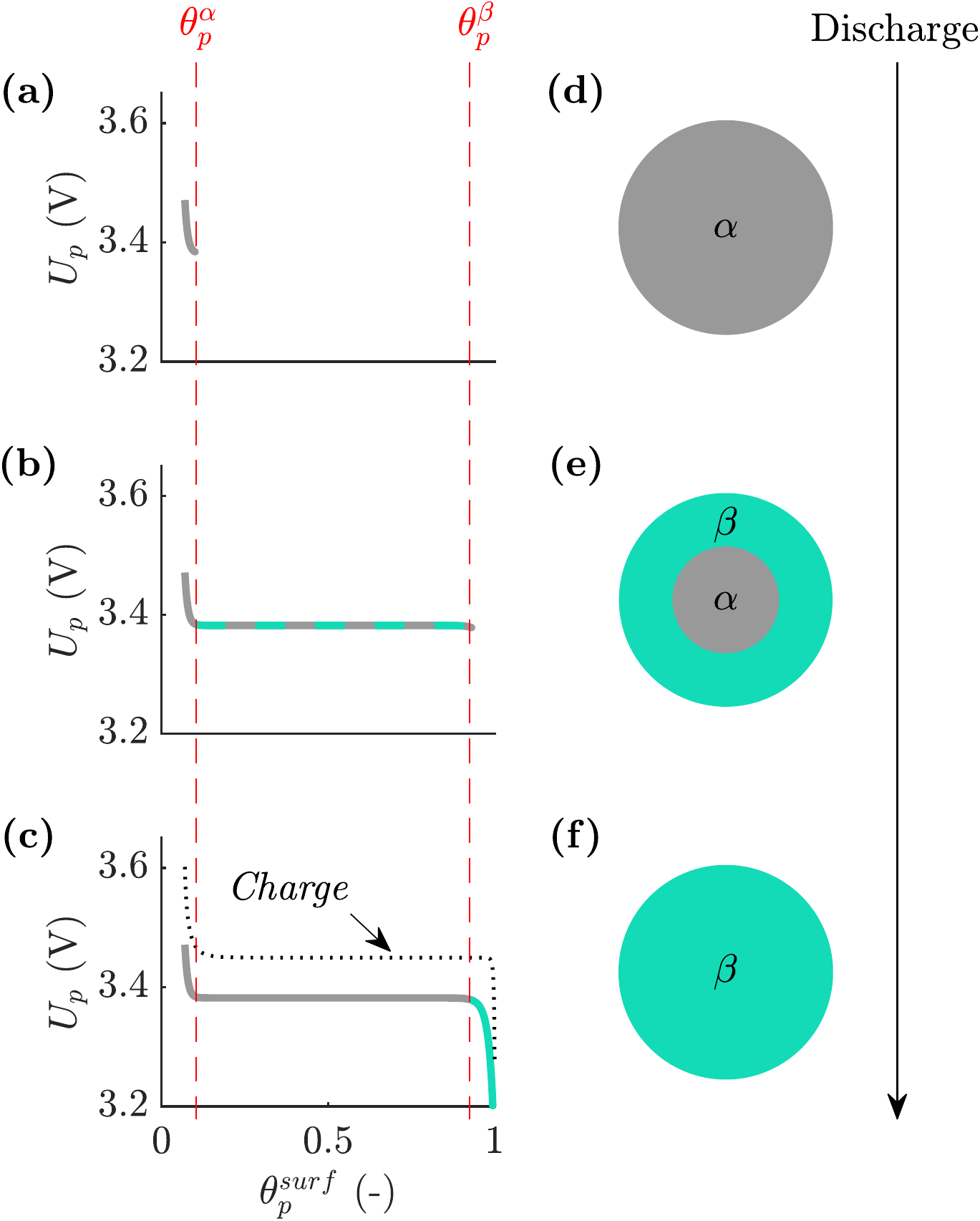}
\caption{ Figures (a), (b), and (c) show the behavior of the positive particle OCP during discharge (the black dotted line represents the charge OCP).  During the one-phase regions (a) and (c), the potential decreases, whereas in (b), the coexistence of two phases leads to a flat OCP.  Figures (d), (e), and (f) show the core-shell representation of the particle used to describe the one-phase regions ((d) and (f)) and the transition from the $\alpha$- to $\beta$-phase (e).}
\label{fig:bsfc}
\end{figure}
To understand the intercalation and deintercalation phenomena in LFP batteries,  the positive particle OCP ($U_p$) is analyzed during discharge (Fig.  \ref{fig:bsfc} (a), (b),  and (c)).  First,  lithium intercalates into the positive electrode and the $\alpha$-phase is formed.  This corresponds to Fig.  \ref{fig:bsfc}(a), with a rapid decrease of the OCP.  As intercalation continues,  lithium concentration increases and, once the normalized concentration $\theta_p^\alpha$ is reached,  the formation of the $\beta$-phase starts.  In this condition,  the $\alpha$-phase transitions into the $\beta$-phase and, while these two phases coexist (Fig.  \ref{fig:bsfc}(b)),  the OCP remains constant.  The transition ends at the normalized concentration $\theta_p^\beta$ and, after this point, the positive particle is all at $\beta$-phase and  the OCP decreases until the end of discharge (Fig. \ref{fig:bsfc}(c)).   During charge,  the process is reversed\,--\,from (c) to (a)\,--\,and a different OCP is followed.  As shown in \cite{dreyer2010thermodynamic},  during charge the surface of the positive electrode is in $\beta$-phase,  which produces  the $Charge$ OCP shown in Fig. \ref{fig:bsfc}(c).  

\subsection{Model development}\label{sec:model_dev}
To model the positive particle phase transition described in Section \ref{sec:physics},  the ESPM developed in \cite{allam2018interconnected} is used as a starting point.  In the ESPM,  both the positive and negative spherical particles are modeled in terms of mass and charge transport in the electrolyte phase (Eqs. \eqref{goveq:eq2} and \eqref{goveq:eq3}),  and mass transport in the solid phase (Eq. \eqref{goveq:eq4}).  ESPM is limited for LFP applications because it does not allow for the description of the positive particle transition from $\alpha$- to $\beta$-phase.  
In this work, the positive particle dynamics in discharge are modeled to reproduce the phases depicted in Fig.~\ref{fig:bsfc} (d), (e), and (f).
When the discharge process starts,  the particle is in $\alpha$-phase (d).  In this scenario,  the positive particle solid phase concentration is modeled via ESPM,  namely using Eq. \eqref{goveq:eq4} (and boundary conditions \eqref{goveq:eq4b} and \eqref{goveq:eq5}).    
While the positive electrode is being lithiated,  the OCP decreases until the normalized concentration reaches $\theta_p^\alpha$ and, after this point, the formation of the $\beta$-phase starts.  Two phases coexist inside the particle at this point which are modeled using the core-shell paradigm.  As shown in  Fig. \ref{fig:bsfc}(e),  the core of the cell is assumed to be at one phase ($\alpha$) and then as intercalation evolves,  covered by the surrounding shell phase ($\beta$).  The core is assumed to be at a constant and uniform concentration $c_{s,p}^\alpha = \theta_p^\alpha\cdot c_{s,p}^{max}$ and subject to a shrinking process which replaces the $\alpha$- with $\beta$-phase.  This process is modeled via the following mass balance:
\begin{equation}
\mathrm{sign}(I)(c_{s,p}^\alpha-c_{s,p}^\beta)\frac{d r_p}{d t} = D_{s,p}\frac{\partial c_{s,p}}{\partial r}\bigg|_{r = r_p}
\label{eq:massbal}
\end{equation}
describing the motion of the interface (or boundary) $r_p$ between the $\alpha$- and $\beta$-phase while assuming $d r_p/d t$ to be function of the concentration gradient $\partial c_{s,p}/\partial r$ only.  In Eq. \eqref{eq:massbal},  $c_{s,p}^\alpha$ and $c_{s,p}^\beta$ are constants defined with respect to $ \theta_p^\alpha$ and $ \theta_p^\beta$,  respectively (see Eq. \eqref{eq:currconv_1}),  $c_{s,p}$ is the solid phase concentration,  $D_{s,p}$ is the solid phase diffusion coefficient,  and $r$ is the radial coordinate (i.e., the coordinate along the radius of the particle).  The  term $\mathrm{sign}(I)$ accounts for the fact that, during discharge,  $\alpha$-phase transitions to $\beta$-phase and that the opposite happens during charge.  The moving boundary $r_p$ corresponds to the distance between the center of the particle and the interface between $\alpha$ and $\beta$.  To complete the model,  the following boundary and initial conditions are introduced:
\begin{equation}
c_{s,p}\big|_{r=r_p} = c_{s,p}^{\beta},\quad c_{s,p}\big|_{t=\bar{t}} = c_{s,p}^{\alpha}, \quad r_p\big|_{t=\bar{t}} = R_p-\epsilon
\label{eq:ictrans}
\end{equation}
The first equality enforces the concentration at the interface $r_p$ to be always equal to $c_{s,p}^\beta$.  The second and third equalities are initial conditions for the core-shell model when, at the time instant $\bar{t}$,  the system moves from Fig.  \ref{fig:bsfc}(d) to (e).  These conditions enforce the core to be at a uniform and constant concentration $c_{s,p}^{\alpha}$ and $r_p$ to be equal to $R_p-\epsilon$ (with $\epsilon$ small enough). In the shell region,  $c_{s,p}^\beta$ corresponds to the initial condition and, for $r> r_p$,  the solid phase concentration rises according to Eq. \eqref{goveq:eq4}.  Finally,  once the core is completely consumed,  the particle is fully in  $\beta$-phase and intercalation occurs until the end of the discharge process,  as shown in Fig. \ref{fig:bsfc}(f).  During charge,  the opposite process occurs.  In this scenario,  the core is in $\beta$-phase and shrinks until the whole particle is in $\alpha$-phase.  General equations for charge and discharge are provided in \eqref{goveq:eq6},  \eqref{goveq:eq7},  \eqref{eq:currconv_2},  and \eqref{eq:currconv_3}.  Governing equations are summarized in Table \ref{table:ESPM_table_1}.  Additional equations for the core-shell ESPM\,--\,transport parameters, active area, porosity,  cell voltage, overpotential, and $SOC$\,--\,are summarized in Table \ref{table:ESPM_table_2}.

\begin{table}[!tb]
	\caption{Governing equations of the core-shell ESPM model.}\label{table:ESPM_table_1}
	\Centering	
	\resizebox{\columnwidth}{!}{
	\scriptsize{				
		\begin{tabular}{l}
			\hline\hline \\ [-3mm]
           \parbox{8cm}{
			    \begin{equation}
			    \label{goveq:eq1}
			    \begin{split}
			    &J_n = \frac{I}{A_{cell}FL_n},\quad J_p = \frac{-I}{A_{cell}FL_p}, \quad J_s = 0\hspace{7em}\\
			    \end{split}
				\end{equation}}\\
             \textbf{Mass transport in the electrolyte phase, } $i\in\mathcal{M}$ \\
			\parbox{8cm}{
			    \begin{equation}
			     \label{goveq:eq2}
			    \begin{split}
			    & \varepsilon_i\frac{\partial c}{\partial t} = \frac{\partial}{\partial x}\left(D_{eff,i}(c,T)\frac{\partial c}{\partial x}\right) + (1-t_+)J_i\hspace{9em}\\
			    \end{split}
				\end{equation}}\\
		    \textbf{Charge transport in the electrolyte phase, } $i\in\mathcal{M}$ \\ 
		    \parbox{8cm}{
			    \begin{equation}
			     \label{goveq:eq3}
			    \begin{split}
			    &\kappa_{eff,i}(c)\frac{\partial}{\partial x}\left(\frac{\partial \phi_e}{\partial x}\right) - \frac{2RT\kappa_{eff,i}(c)v(c,T)}{F}\frac{\partial^2\ln(c)}{\partial x^2}+FJ_i=0\\
			    \end{split}
				\end{equation}} \\
		    \textbf{Mass transport in the solid phase,} $i\in\hat{\mathcal{M}}$ \\ 
		    \parbox{8cm}{
			    \begin{equation}
			    \label{goveq:eq4} 
			    \frac{\partial c_{s,i}}{\partial t} = D_{s,i}\frac{\partial^2c_{s,i}}{\partial r^2}+\frac{2D_{s,i}}{r}\frac{\partial c_{s,i}}{\partial r}\hspace{13em}
				\end{equation}}\\ 
		      \parbox{8cm}{
			    \begin{equation}
			    \label{goveq:eq4b} 
			    \frac{\partial c_{s,i}}{\partial r}\bigg |_{r=0}=0 \hspace{23em}
				\end{equation}}\\
		     \parbox{8cm}{
			    \begin{equation}
			    \label{goveq:eq5} 
			    \frac{\partial c_{s,p}}{\partial r}\bigg\vert_{r = R_p} \hspace{-1.5em}= \frac{I}{D_{s,p}a_{p}A_{cell}FL_p},\ \ \frac{\partial c_{s,n}}{\partial r}\bigg\vert_{r = R_n} \hspace{-1.5em}= \frac{-I}{D_{s,n}a_{n}A_{cell}FL_n}
				\end{equation}}\\
			\parbox{8cm}{
			    \begin{subequations}
			    \begin{empheq}[left=\rotatebox{90}{\hspace{-1.8em}\text{Core-shell}} \empheqlbrace]{align}
                 & \mathrm{sign}(I)(c_{s,p}^\alpha-c_{s,p}^\beta)\frac{d r_p}{d t} = D_{s,p}\frac{\partial c_{s,p}}{\partial r}\bigg|_{r = r_p}  \label{goveq:eq6}\\
			    &c_{s,p}\big|_{r=r_p} = \mathrm{g}(I), \ \ c_{s,p}\big|_{t=\bar{t}} = \mathrm{ic}_k, \ \ r_p\big|_{t=\bar{t}} = R_p-\epsilon \hspace{0em}\label{goveq:eq7}
			    \end{empheq}
				\end{subequations}}\\ [3mm]
			\hline\hline 
		\end{tabular}}}
\end{table}

\begin{table}[!tb]
	\renewcommand{\arraystretch}{1.5}
	\caption{Additional equations for the core-shell ESPM model.}\label{table:ESPM_table_2}
	\centering
	\resizebox{\columnwidth}{!}{					
		\begin{tabular}{l}
			\hline\hline \\ [-5mm]
		     \multicolumn{1}{l}{\textbf{Current convention}}  \\ [-2mm]
			\parbox{10cm}{
			   \begin{flalign} \label{eq:currconv_1}
				& \quad\begin{cases}
				I > 0, \ \ \mathrm{discharge}\\
				I = 0, \\
				I < 0, \ \ \mathrm{charge}\\
				\end{cases} &
				\end{flalign}}\\[-2mm]
				
			 \multicolumn{1}{l}{\textbf{Concentration at the moving boundary}}  \\ [-2mm]
		    	\parbox{10cm}{
			   \begin{flalign} \label{eq:currconv_2}
				&  \quad\mathrm{g}(I) = \begin{cases}
				c_{s,p}^\beta = \theta_p^\beta\cdot c_{s,p}^{max}, \ \ \text{if}\ I > 0\\
	             c_{s,p}^\alpha  = \theta_p^\alpha\cdot c_{s,p}^{max}, \ \text{if}\ I < 0\\
				0,\ \ \text{otherwise}\\
				\end{cases}  &
				\end{flalign}}\\[-2mm]
				
			\multicolumn{1}{l}{\textbf{Core initial condition}}  \\ [-2mm]
			\parbox{10cm}{
			    \begin{flalign} \label{eq:currconv_3}
				& \quad\mathrm{ic}_k = \begin{cases}
				c_{s,p}^\alpha, \ k = \mathrm{discharge}\\
				c_{s,p}^\beta,\ k = \mathrm{charge} \\
				\end{cases} &
				\end{flalign}}\\[-2mm]
				
			\multicolumn{1}{l}{\textbf{Diffusivity and conductivity}}  \\ [-2mm]
			\parbox{10cm}{
			    \begin{flalign} \label{eq:diff_1}
				& \quad D_{eff,i}(c,T) = D(c,T)\cdot \varepsilon_i^{brugg},\ \ i \in\mathcal{M} &
				\end{flalign}
				\begin{flalign*}
				& \quad\quad\rightarrow D(c,T) = 0.0001\cdot 10^{\left(-4.51-\frac{59.22}{T-(206.25+10c/1000)}\right)c/1000} &
				\end{flalign*}} \\[-2mm]
			\parbox{10cm}{
			    \begin{flalign} \label{eq:diff_2}
				& \quad \kappa_{eff,i}(c) = \kappa(c)\cdot \varepsilon_i^{brugg}, \ \ i \in\mathcal{M} &
				\end{flalign}
				\begin{flalign*}
				& \quad\quad\rightarrow \kappa(c) = \left(\frac{c^{avg}/1000}{1.05}\right)^{0.68} 
				 \mathrm{exp}[-0.1(c^{avg}/1000-1.05)^2+\\ &\hspace{5.6em}- 0.56\left(c^{avg}/1000-1.05\right)] &
				\end{flalign*}} \\[-1mm]
				
		     \multicolumn{1}{l}{\textbf{Active area}}  \\ [-2mm]
			\parbox{10cm}{
			    \begin{flalign} \label{eq:act_area}
				& \quad a_i=\frac{3}{R_i}\nu_i,\ \ i \in\hat{\mathcal{M}} &
				\end{flalign}} \\[-1mm]
				
			\multicolumn{1}{l}{\textbf{Porosity}}  \\ [-2mm]
			\parbox{10cm}{
			    \begin{flalign} \label{eq:poro_1}
				& \quad \varepsilon_{i} = 1-\nu_i-\nu_{i,filler},\ \ i \in\hat{\mathcal{M}} &
				\end{flalign}} \\[-1mm]
				
			\multicolumn{1}{l}{\textbf{Cell voltage}}  \\ [-2mm]
			\parbox{10cm}{
			    \begin{flalign} \label{eq:cell_volt_3}
				& \quad  \Phi_{s,i} =U_i(\theta_i^{surf}) + \eta_i,\ \ i \in\hat{\mathcal{M}} &&
				\end{flalign}} \\[-2mm]
		     \parbox{10cm}{
			    \begin{flalign} \label{eq:cell_volt_4}
				& \quad  \Delta\Phi_e = \frac{2RTv(c,T)}{F}\ln\left(\frac{c(L)}{c(0)}\right), \ \mathrm{with}\ L = L_n+L_s+L_p & 
				\end{flalign} \vspace{-1.8em}
				\begin{flalign*}
		         & \quad\quad\rightarrow v(c,T) = 0.601 - 0.24(c^{avg}/1000)^{1/2} + \\ &\quad\hspace{4.5em} + 0.982\left[1-0.0052(T-293)\right](c^{avg}/1000)^{3/2}\ \text{\cite{tanim2015temperature}} &
				\end{flalign*}} \\[-2mm]
			\parbox{10cm}{
			    \begin{flalign} \label{eq:cell_volt_1}
				& \quad V = \Phi_{s,p}-\Phi_{s,n}+\Delta\Phi_{e} - I(R_l+R_{el}) &
				\end{flalign}} \\[-2mm]
			\parbox{10cm}{
			    \begin{flalign} \label{eq:cell_volt_2}
				& \quad R_{el} = \frac{1}{2A_{cell}} \left(\frac{L_n}{\kappa_{eff,n}(c)}+\frac{2L_s}{\kappa_{eff,s}(c)}+\frac{L_p}{\kappa_{eff,p}(c)}\right) &
				\end{flalign}} \\[-2mm]
	         \parbox{10cm}{
			    \begin{flalign} \label{eq:cell_volt_5}
				& \begin{cases}
				\begin{split} U_p &= 3.382 -0.2955 \exp{\left[-44.99(1-\theta_{p}^{surf})^{0.8707}\right]} +\\
    &+ 10^{-20.71} \exp{\left[14.17 (1-\theta_{p}^{surf})^{8.128}\right]} +\\[1mm]
   &+ 10^{-40.82} \exp{\left[100(1-\theta_{p}^{surf})^{1.213}\right]}\end{split}, \quad \mathrm{discharge}\\
				\begin{split}     U_p &= 3.442 -0.1774 \exp{\left[-127.7(1-\theta_{p}^{surf})^{0.7921}\right]} +\\
    &+ 10^{-2.123}\exp{\left[16.56 (1-\theta_{p}^{surf})^{24.08}\right]}+ \\[1mm]
    &+ 10^{-10.29}\exp{\left[99.91(1-\theta_{p}^{surf})^{22.17}\right]}\end{split},\quad \mathrm{charge} \end{cases}&
				\end{flalign}} \\[-2mm]
				\parbox{10cm}{
			    \begin{flalign} \label{eq:cell_volt_6}
                 &\quad \theta_{n}^{surf}  = c_{s,n}/c_{s,n}^{max},\quad \theta_{p}^{surf}  = c_{s,p}/c_{s,p}^{max} & 
				\end{flalign}} \\[-1mm]
				
			\multicolumn{1}{l}{\textbf{Electrochemical overpotential}}  \\ [-2mm]
			\parbox{10cm}{
			    \begin{flalign} \label{eq:overp_1}
				& \quad \eta_i =  \frac{RT}{0.5F}\sinh^{-1}\left(\frac{I}{2A_{cell}\ a_{i}\ L_i\ i_{0,i}}\right),\ \ i \in\hat{\mathcal{M}} &
				\end{flalign}} \\[-2mm]
			\parbox{10cm}{
			    \begin{flalign} \label{eq:overp_2}
				& \quad i_{0,i} = k_iF\sqrt{c^{avg}c_{s,i}^{surf}\left(c_{s,i}^{max}-c_{s,i}^{surf}\right)},\ \ i \in\hat{\mathcal{M}} &
				\end{flalign}} \\[-1mm]				
		
		    		     \multicolumn{1}{l}{\textbf{State of charge}}  \\ [-2mm]
			\parbox{10cm}{
			    \begin{flalign} \label{eq:soc_1}
                  \begin{split}
                  & \quad SOC_n = \frac{\theta_n^{bulk}-\theta_{n,0\%}}{\theta_{n,100\%}-\theta_{n,0\%}},\ SOC_p = \frac{\theta_{p,0\%}-\theta_p^{bulk}}{\theta_{p,0\%}-\theta_{p,100\%}}\hspace{15em}\\ 
                  & \quad \mathrm{Negative\ particle}\\
                  & \quad\quad\quad \theta_{n}^{bulk}  = \frac{3}{c_{s,n}^{max}R_n^3}\int_{0}^{R_n}c_{s,n}r^2dr\\
                  & \quad \mathrm{Positive\ particle}\\
                  & \quad\quad\quad \theta_{p}^{bulk}  = \frac{3}{c_{s,p}^{max}R_p^3}\int_{0}^{R_p}c_{s,p}r^2dr
                  \end{split}
				\end{flalign}} \\[-1mm]
			
			\hline\hline
		\end{tabular}}
		\
		\renewcommand{\arraystretch}{1.4}
\end{table}

\section{Numerical solution approach}\label{sec:numericalsol}
Eqs. in Table \ref{table:ESPM_table_1} constitute a system of coupled PDEs.  In this section,  PDEs are discretized using the finite difference method (FDM),  for mass transport in the solid phase,  and finite volume method (FVM),   for mass transport in the electrolyte phase. The system of ODEs obtained  from the numerical discretization is converted into a convenient state-space representation and solved relying on numerical solvers such as \texttt{ode15s}. 
In the following, Eqs. \eqref{goveq:eq4},  \eqref{goveq:eq5},  \eqref{goveq:eq6},  and \eqref{goveq:eq7}  are analyzed and discretized using FDM.  Further details on the solution of the remaining governing equations for the electrolyte and negative particle can be found in  \cite{weaver2020novel}. The reader is referred to \cite{pozzato2021modeling} for full explanation of notation and symbols used in this paper. 

\subsection{Coordinate system transformation}
Starting from the positive particle core-shell model described in Section \ref{sec:governingeq}, the transformation proposed in \cite{srinivasan2004discharge} is used to move from the radial coordinate system to the normalized coordinate $\chi$, $\chi = \frac{r - r_p}{R_p - r_p} \in [0,1]$, where $r$ represents a given radial position in the particle and $r_p$ the position of the moving boundary.  This transformation allows to remap the discretization of the shell region from $[r_p,R_p]$ to $[0,1]$,  making the domain stationary while the boundary is moving.  In the initial condition $r_p\big|_{t=\bar{t}} = R_p-\epsilon$ (Eq.  \eqref{eq:ictrans}),  the small enough $\epsilon$ avoids the rise of singularities during transitions from one-phase to two-phase condition. According to \cite{li2015modeling}, the left hand side of the positive particle diffusion Eq. \eqref{goveq:eq4} can be transformed from $r$  to $\chi$ domain as follows:
\begin{equation}
\left(\frac{\partial c_{s,p}}{\partial t}\right)_r = \frac{\partial c_{s,p}}{\partial \chi}\frac{\partial \chi}{\partial t} + \left(\frac{\partial c_{s,p}}{\partial t}\right)_\chi
\label{eq:term_1}
\end{equation}
From the following relationships $\frac{\partial\chi}{\partial t} = \frac{\partial \chi}{\partial r_p}\frac{\partial r_p}{\partial t},\quad \frac{\partial \chi}{\partial r_p} = \frac{\chi-1}{R_p-r_p}$, Eq. \eqref{eq:term_1} is rewritten as:
\begin{equation}
\left(\frac{\partial c_{s,p}}{\partial t}\right)_r = \frac{\partial c_{s,p}}{\partial \chi} \left(\frac{\chi-1}{R_p-r_p}\right)\frac{\partial r_p}{\partial t} + \left(\frac{\partial c_{s,p}}{\partial t}\right)_\chi
\label{eq:part_1}
\end{equation}
The change of coordinate system for the right hand side terms of Eq. \eqref{goveq:eq4} is obtained introducing the following relationships:
\begin{equation}
\begin{split}
&\frac{\partial^2 c_{s,p}}{\partial r^2} = \frac{\partial}{\partial r}\left(\frac{\partial c_{s,p}}{\partial r}\right) = \frac{\partial}{\partial r}\left(\frac{\partial c_{s,p}}{\partial \chi}\frac{\partial\chi}{\partial r}\right)\\
 \frac{\partial c_{s,p}}{\partial r}&=\frac{\partial c_{s,p}}{\partial \chi}\frac{\partial\chi}{\partial r}, \quad\frac{\partial}{\partial r} = \frac{\partial\chi}{\partial r}\frac{\partial}{\partial\chi}, \quad \frac{\partial\chi}{\partial r} = \frac{1}{R_p-r_p}
\end{split}
\label{eq:changecoord_2}
\end{equation}
Therefore, relying on Eq. \eqref{eq:changecoord_2}, the terms on the right hand side are reformulated as follows:
\begin{equation}
D_{s,p}\frac{\partial^2 c_{s,p}}{\partial r^2} = \frac{\partial^2 c_{s,p}}{\partial \chi^2}\frac{D_{s,p}}{(R_p-r_p)^2}
\label{eq:part_2}
\end{equation}
\begin{equation}
\frac{2 D_{s,p}}{r}\frac{\partial c_{s,p}}{\partial r} = \frac{\partial c_{s,p}}{\partial\chi}\left[\frac{2D_{s,p}}{r(R_p-r_p)}\right]
\label{eq:part_3}
\end{equation}
From Eqs. \eqref{eq:part_1},  \eqref{eq:part_2},  and \eqref{eq:part_3},  the solid phase mass transport in the positive particle (Eq. \eqref{goveq:eq4}) is rewritten as:
\begin{equation}
\begin{split}
\frac{\partial c_{s,p}}{\partial t} &= \frac{\partial^2 c_{s,p}}{\partial\chi^2}\left[\frac{D_{s,p}}{(R_p-r_p)^2}\right]+\\&+\frac{\partial c_{s,p}}{\partial\chi}\left[\frac{2D_{s,p}}{r(R_p-r_p)}\right]-\frac{\partial c_{s,p}}{\partial\chi}\frac{\partial r_p}{\partial t}\left[\frac{\chi-1}{R_p-r_p}\right]\\
\end{split}
\label{eq:final_1}
\end{equation}

\noindent Finally,  starting from Eq. \eqref{eq:changecoord_2},  boundary conditions \eqref{goveq:eq5}, \eqref{goveq:eq7},  and the mass balance \eqref{goveq:eq6} are also rewritten in terms of the $\chi$ coordinate:  
\begin{equation}
\frac{\partial c_{s,p}}{\partial \chi}\bigg\vert_{\chi=1} = \frac{I(R_p-r_p)}{D_{s,p}a_{p}A_{cell}FL_p}
\label{eq:final_2}
\end{equation}
\begin{equation}
c_{s,p}\big|_{\chi=0} = \mathrm{g}(I) \quad c_{s,p}\big|_{t=\bar{t}} = \mathrm{ic}_k
\label{eq:final_3}
\end{equation}
\begin{equation}
\mathrm{sign}(I)(c_{s,p}^\alpha - c_{s,p}^\beta)(R_p-r_p)\frac{d r_p}{d t} = D_{s,p}\frac{\partial c_{s,p}}{\partial \chi}\bigg|_{\chi=0} 
\label{eq:final_4}
\end{equation}
where $\chi = 0$ coincides with the moving boundary $r_p$ and $\chi=1$ corresponds to the surface or, in other words,  $R_p$.

\subsection{Discretization}
The core-shell model, described by Eqs. \eqref{eq:final_1},  \eqref{eq:final_2}, \eqref{eq:final_3},  and \eqref{eq:final_4},   is discretized into $N_r$ nodes.  The right-sided and central finite difference schemes are used for the first and second derivative approximations: $
\frac{\partial u}{\partial \chi} \big\vert_{\chi_l} \approx \frac{u_{l+1} -  u_l}{\Delta_\chi},\;\; \frac{\partial^2 u}{\partial \chi^2} \big\vert_{\chi_l} \approx \frac{u_{l+1} -  2u_l + u_{l-1}}{\Delta_\chi^2}$, where $u$ is a mute variable and $l$ defines the index of the discretization point $\chi_l$ such that  $\chi_l = \frac{r_l-r_p}{R_p-r_p},\ \Delta_\chi = \chi_l - \chi_{l-1}$
where  $r_l$,  the radial position along the shell region,  takes the following form $
r_l=r_p+l\Delta_r, 	\ \Delta_r = \frac{R_p-r_p}{N_r-1}$.
\subsubsection{Discretization of mass balance}
First, Eq. \eqref{eq:final_4} is rewritten as:
\begin{equation}
\frac{d r_p}{d t} = \frac{\mathrm{sign}(I)D_{s,p}}{(c_{s,p}^\alpha- c _{s,p}^\beta)(R_p-r_p)}\frac{\partial c_{s,p}}{\partial\chi}\bigg\vert_{0}
\end{equation}
Approximating the term $\frac{\partial c_{s,p}}{\partial\chi}\big\vert_{0}$ as
$\frac{c_{s,p}}{\partial\chi}\big\vert_{0}\approx \frac{c_{s,p_1}-c_{s,p_0}}{\Delta_\chi}$
and knowing from Eq.~\eqref{eq:discr_1} (introduced in the next paragraph) that $c_{s,p_{0}} = \mathrm{g}(I)$ , the discretized version of the mass balance is obtained:
\begin{equation}
\begin{split}
\frac{d r_p}{d t} = \frac{M_1}{\Delta_\chi}(c_{s,p_1}-\mathrm{g}(I)),
M_1 = \frac{\mathrm{sign}(I)D_{s,p}}{(c_{s,p}^\alpha - c_{s,p}^\beta)(R_p-r_p)}
\end{split}
\label{eq:discr_3}
\end{equation}

\subsubsection{Discretization of boundary conditions}
The moving boundary condition at $\chi = 0$ (Eq. \eqref{eq:final_3}) takes the following form:
\begin{equation}
c_{s,p_{0}} = \mathrm{g}(I) \rightarrow \frac{\partial c_{s,p_0}}{\partial t} = 0
\label{eq:discr_1}
\end{equation}
The fixed boundary condition \eqref{eq:final_2} is rewritten as:
\begin{equation}
\frac{\partial c_{s,p}}{\partial \chi}\bigg\vert_{N_r-1} = \frac{I(R_p-r_p)}{D_{s,p}a_{p}A_{cell}FL_p}
\label{eq:fb_disc}
\end{equation}
Approximating the left hand side of Eq. \eqref{eq:fb_disc} as
$\frac{\partial c_{s,p}}{\partial \chi}\big\vert_{N_r-1} \approx \frac{c_{s,p_{N_r}}-c_{s,p_{N_r-1}}}{\Delta_\chi}$, 
the following is obtained:
\begin{equation}
\begin{split}
c_{s,p_{N_r}} = c_{s,p_{N_r-1}} + M_2I, \; M_2 = \frac{(R_p-r_p)\Delta_\chi}{D_{s,p}a_pA_{cell}FL_p}
\end{split}
\label{eq:discr_2}
\end{equation}
\subsubsection{Discretization of the solid phase mass transport}
Eq. \eqref{eq:final_1} is discretized using both the right-sided and central finite difference schemes. For $l \in [1,N_r-2]$, the following expression is obtained:
\begin{equation}
\resizebox{\columnwidth}{!}{$
\begin{split}
\frac{\partial c_{s,p}}{\partial t}\bigg\vert_{l} &= \frac{M_3}{\Delta_\chi^2}(c_{s,p_{l+1}}-2c_{s,p_l} +c_{s,p_{l-1}}) +\frac{M_4}{\Delta_\chi}(c_{s,p_{l+1}}-c_{s,p_l})
\end{split}$}
\label{eq:discr_4}
\end{equation}
with $M_3$ and $M_4$ defined as $M_3 = \frac{D_{s,p}}{(R_p-r_p)^2}\vspace{0.2em},\;M_4 = \frac{2D_{s,p}}{[\chi_l(R_p-r_p)+r_p](R_p-r_p)}-\frac{\chi_l-1}{R_p-r_p}\frac{M_1}{\Delta_\chi}(c_{s,p_1}-\mathrm{g}(I))\vspace{0.5em}$.
At $l=N_r-1$,  the discretized solid phase mass transport equation takes the following form $ \frac{\partial c_{s,p}}{\partial t}\big\vert_{N_r-1}= \frac{M_3}{\Delta_\chi^2}(c_{s,p_{N_r}}-2c_{s,p_{N_r-1}} +c_{s,p_{N_r-2}})+\frac{M_4}{\Delta_\chi}(c_{s,p_{N_r}}-c_{s,p_{N_r-1}})$
which, using Eq. \eqref{eq:discr_2}, is rewritten as:
\begin{equation}
\begin{split}
\frac{\partial c_{s,p}}{\partial t}\bigg\vert_{N_r-1} &= \frac{M_3}{\Delta_\chi^2}(M_2I-c_{s,p_{N_r-1}} +c_{s,p_{N_r-2}})+\frac{M_2M_4}{\Delta_\chi}I
\end{split}
\label{eq:discr_5}
\end{equation}
From  Eq. \eqref{eq:discr_1}, at $l=0$ the time derivative $\frac{\partial c_{s,p_0}}{\partial t} = 0$.

\subsection{State-space representation}
Eqs. \eqref{eq:discr_3},  \eqref{eq:discr_1}, \eqref{eq:discr_4}, and \eqref{eq:discr_5} constitute a system of coupled ODEs  which can be conveniently rewritten in state-space form.  
Given $r_p$ and the vector of discretized solid phase concentration states: $
\mathbf{c}_{s,p} = [c_{s,p_1}\ c_{s,p_2}\ \dots\ c_{s,p_{N_r-2}}\ c_{s,p_{N_r-1}}]^T \in \mathbb{R}^{(N_r-1)\times 1}$,
the following state vector is defined $
\mathbf{x} = \begin{bmatrix} r_p\\ \mathbf{c}_{s,p}\end{bmatrix}\in \mathbb{R}^{N_r\times 1}$. 
Introducing the variables $\eta_1 = \frac{M_3}{\Delta_\chi^2},\quad \eta_2 = \frac{M_4}{\Delta_\chi},\ \eta_3 = \frac{M_2}{\Delta_\chi}\left(M_4+\frac{M_3}{\Delta_\chi}\right), \ \text{and}\ \eta_4 = \frac{M_1}{\Delta_\chi}$,  
the positive particle state-space representation takes the following form:
\begin{equation}
\dot{\mathbf{x}} = \eta_1\mathbf{A}_1\mathbf{x} + \eta_2\mathbf{A}_2\mathbf{x} + \eta_3\mathbf{B}I+\eta_1\mathbf{G}
\label{eq:sscore}
\end{equation}
where matrices $\mathbf{A}_1$ and $\mathbf{A}_2$ are given by:
\begin{equation}
\scriptsize{\mathbf{A}_1 = 
\begin{bmatrix}
0 & \eta_4/\eta_1  & 0 & 0 & 0 & \dots & 0  \\
0 & -2 & 1 & 0 & 0 & \dots & 0  \\
0 &1 & -2 & 1 & 0 & \dots & 0 \\
0 &0 & 1 & -2 & 1 & \dots & 0 \\
0 &0 & 0 & 1 & -2 & \dots & 0 \\
\vdots & \vdots & \vdots & \vdots & \vdots & \ddots & \vdots\\
0& 0 & 0 & 0 & 0 & \dots &  -1\\
\end{bmatrix}_{N_r\times N_r}}
\end{equation}
\begin{equation}
\scriptsize{\mathbf{A}_2 = 
\begin{bmatrix}
0 & 0 & 0 & 0 & 0 & \dots & 0 \\
0 & -1 & 1 & 0 & 0 & \dots & 0 \\
0 & 0 & -1 & 1 & 0 & \dots & 0 \\
0 &0 & 0 & -1 & 1 & \dots & 0 \\
0 &0 & 0 &  0 & -1 & \dots & 0 \\
\vdots & \vdots & \vdots & \vdots & \vdots & \ddots & \vdots \\
0 &0 & 0 & 0 & 0 & \dots & 0 \\
\end{bmatrix}_{N_r\times N_r}}
\end{equation}
and  vectors $\mathbf{B}$ and $\mathbf{G}$ are defined as:
\begin{equation}
\scriptsize{\mathbf{B} = 
\begin{bmatrix}
0\\
0\\
0\\
\vdots\\
0\\
1\\
\end{bmatrix}_{N_r\times 1}, \quad \mathbf{G} = 
\begin{bmatrix}
-\eta_4/\eta_1\mathrm{g}(I)\\
\mathrm{g}(I)\\
0\\ 
\vdots\\
0\\
0\\
\end{bmatrix}_{N_r\times 1}}
\end{equation}

\section{Parameters identification}\label{sec:paramid}
Model parameters are identified using voltage \textit{vs} capacity data for a $Q_{nom} = 49\mathrm{Ah}$ LFP pouch cell charged and discharged at C/12 constant current (CC),  at 25$^\circ$C.  Using the particle swarm optimization (PSO) algorithm\footnote{\textsc{Matlab} \textit{particleswarm} function: \url{https://www.mathworks.com/help/gads/particleswarm.html}}, the following parameter vector is identified  $\Theta = [ R_n, R_p,A_{cell},D_{s,n},D_{s,p},
\theta_{n,100\%},\theta_{n,0\%},\theta_{p,100\%},\theta_{p,0\%},\theta_p^{\alpha}$,\\$\theta_p^{\beta},R_l]$.  An a posteriori correlation analysis of the identified parameters is shown in \cite{pozzato2022}.
For the core-shell modeling framework to be correctly  implemented, the identification of $\theta_p^\alpha$ and $\theta_p^\beta$ is crucial to properly define the transition from one-phase ($\alpha$ or $\beta$) to two-phase ($\alpha$ and $\beta$).
The identification of the model parameters is performed following the framework established in  \cite{allam2020online}, that relies on  minimizing the following multi-objective cost function both in charge and discharge:
\begin{equation}
\scriptsize{\begin{split}
J_k(\Theta) &= w_1 \sqrt{\frac{1}{N} \sum_{j=1}^N\left(\frac{V_{exp}^k(j) - V^k(\Theta;j)}{V_{exp}^k(j)}\right)^2}  \\
&+ w_2 \sqrt{\frac{1}{N} \sum_{j=1}^N(SOC_{exp}^k(j) - SOC_n^k(\Theta;j))^2} \\
&+ w_3 \sqrt{\frac{1}{N} \sum_{j=1}^N(SOC_{exp}^k(j) - SOC_p^k(\Theta;j))^2} 
\end{split}}
\end{equation}
where  $k\in\mathcal{K} = \{\mathrm{charge},\mathrm{discharge}\}$, $N$ is the number of samples,  $SOC_p^k$ and $SOC_n^k$ are the simulated state of charge at the positive and negative electrodes (Eq. \eqref{eq:soc_1}),  $V^k$ is the simulated voltage profile (Eq. \eqref{eq:cell_volt_1}), $V_{exp}^k$ and $SOC_{exp}^k$ are the experimental cell voltage and state of charge from Coulomb counting, respectively. The weights $w_1$,  $w_2$, and $w_3$ are user-defined dimensionless parameters here equal to one.  The parameter vector is assumed to be the same for charge and discharge because the hysteresis is accounted for by the OCP curves in Eq. \eqref{eq:cell_volt_5}.  In doing so, we also speed up identification by keeping the number of identified parameters at a minimum. 
The multi-objective cost function is subject to the following constrains:
\begin{equation}
\begin{split}
&\mathrm{(a)\ } \text{Governing equations}\ \text{(Table \ref{table:ESPM_table_1})}\\
&\mathrm{(b)\ }\theta^\beta_p \leq \theta_{p,0\%}\\
&\mathrm{(c)\ }\theta^\alpha_p \geq \theta_{p,100\%}\\
&\mathrm{(d)\ }\begin{cases}
	r_p^k(\Theta;j) \geq 0,\quad j\in[1,N-1]\\
	r_p^k(\Theta;N) \leq \rho,\quad j = N,\ \rho\in\mathbb{R}^+
	\end{cases}\\
&\mathrm{(e)\ }\underline{Q} \leq Q_i^k(\Theta) \leq \overline{Q},\quad i\in\hat{\mathcal{M}}
\end{split}\vspace{1em}
\end{equation}
where $Q_i^k$ is the charged/discharged capacity, computed as:
\begin{equation}
Q_i^{k}(\Theta) = \frac{\nu_iFL_iA_{cell}c_{s,i}^{max}\left|\theta_{i,100\%}-\theta_{i,0\%}\right|}{3600}
\end{equation}
Inequalities $\mathrm{(b)}$ and $\mathrm{(c)}$ ensure the two-phase region (defined between $\theta_p^\alpha$ and $\theta_p^\beta$) to be contained inside the positive particle stoichiometric window $\theta_{p,0\%}$-$\theta_{p,100\%}$.  Constraints $\mathrm{(d)}$ enforce the moving boundary $r_p$ to be always positive for $j\in[1,N-1]$ and,  for $j = N$, to be lower than a threshold $\rho$.  In this work,  complete charge and discharge profiles are considered. The cell  reaches the one-phase at the end of the charge or discharge process and the moving boundary $r_p$ reaches zero. 
While $\rho$ should be set to zero,  to soften the constraint and ensure numerical stability we select $\rho = 0.001R_p$ as threshold.  As shown in Section \ref{sec:results},  softening this constraint does not affect the model accuracy.  Finally,  the constraint $\mathrm{(e)}$ ensures charge conservation.  Parameters $\overline{Q}$ and $\underline{Q}$ are suitable bounds.

\renewcommand{\arraystretch}{1.3}
\begin{table}[!tb]
	\caption{Identification results.}	
	\centering
	\label{identification_label}
	\scriptsize{\begin{tabular}{l p{3.5em}  p{3.5em} p{5em} l}
	\toprule
	\textbf{Symbol} & \textbf{Lower bound} & \textbf{Upper bound} & \textbf{Identified vector $\boldsymbol{\Theta}$} & \textbf{Unit} \\
	\midrule
	$R_{n}$ & $1{\times}10^{-6}$ & $2{\times}10^{-5}$  & $1.0{\times}10^{-6}$ & $(\mathrm{m})$ \\
	$R_{p}$ & $1{\times}10^{-8}$ & $1{\times}10^{-5}$  & $4.3{\times}10^{-8}$ & $(\mathrm{m})$ \\
	$A_{cell}$ & $1.41$ & $1.73$ & $1.491$ & $(\mathrm{m^2})$ \\		
	$D_{s,n}$ & $1{\times}10^{-15}$ & $1{\times}10^{-10}$ & $6.9{\times}10^{-12}$ & $(\mathrm{m^2/s})$ \\
	$D_{s,p}$ & $1{\times}10^{-18}$ & $1{\times}10^{-11}$ &  $3.1{\times}10^{-17}$ & $(\mathrm{m^2/s})$ \\	
	{$\theta_{n ,100\%}$} &  $\mathrm{0.7}$ & $\mathrm{0.95}$ & $0.835$ & $(\mathrm{-})$ \\
	{$\theta_{n ,0\%}$} &  $1{\times}10^{-4}$ & $\mathrm{0.2}$  &  $0.010$ & $(\mathrm{-})$ \\
	{$\theta_{p ,100\%}$} &  $\mathrm{0.05}$ & $\mathrm{0.15}$ &  $0.070$ & $(\mathrm{-})$ \\
	{$\theta_{p ,0\%}$} &  $\mathrm{0.8}$ & $\mathrm{1}$ &  $0.882$ & $(\mathrm{-})$ \\			
	{$\theta_p^{\alpha}$} &  $\mathrm{0.1}$ & $\mathrm{0.2}$ &  $0.198$ & $(\mathrm{-})$ \\
	{$\theta_p^{\beta}$} &  $\mathrm{0.8}$ & $\mathrm{0.9}$ &  $0.800$ & $(\mathrm{-})$ \\	
	$R_{l}$ & $1{\times}10^{-3}$ & $\mathrm{0.1}$ & $0.001$ & $(\Omega)$ \\
	\midrule
	 \multicolumn{5}{c}{$\boldsymbol{J(\Theta) = \mathbf{0.011}}$ \textbf{(-)}} \\
	\bottomrule
	\end{tabular}}
\end{table}

\section{Results}\label{sec:results}
\begin{figure}[!tb]
\centering 
\subfloat[\textbf{Discharge}]{\includegraphics[width=0.85\columnwidth]{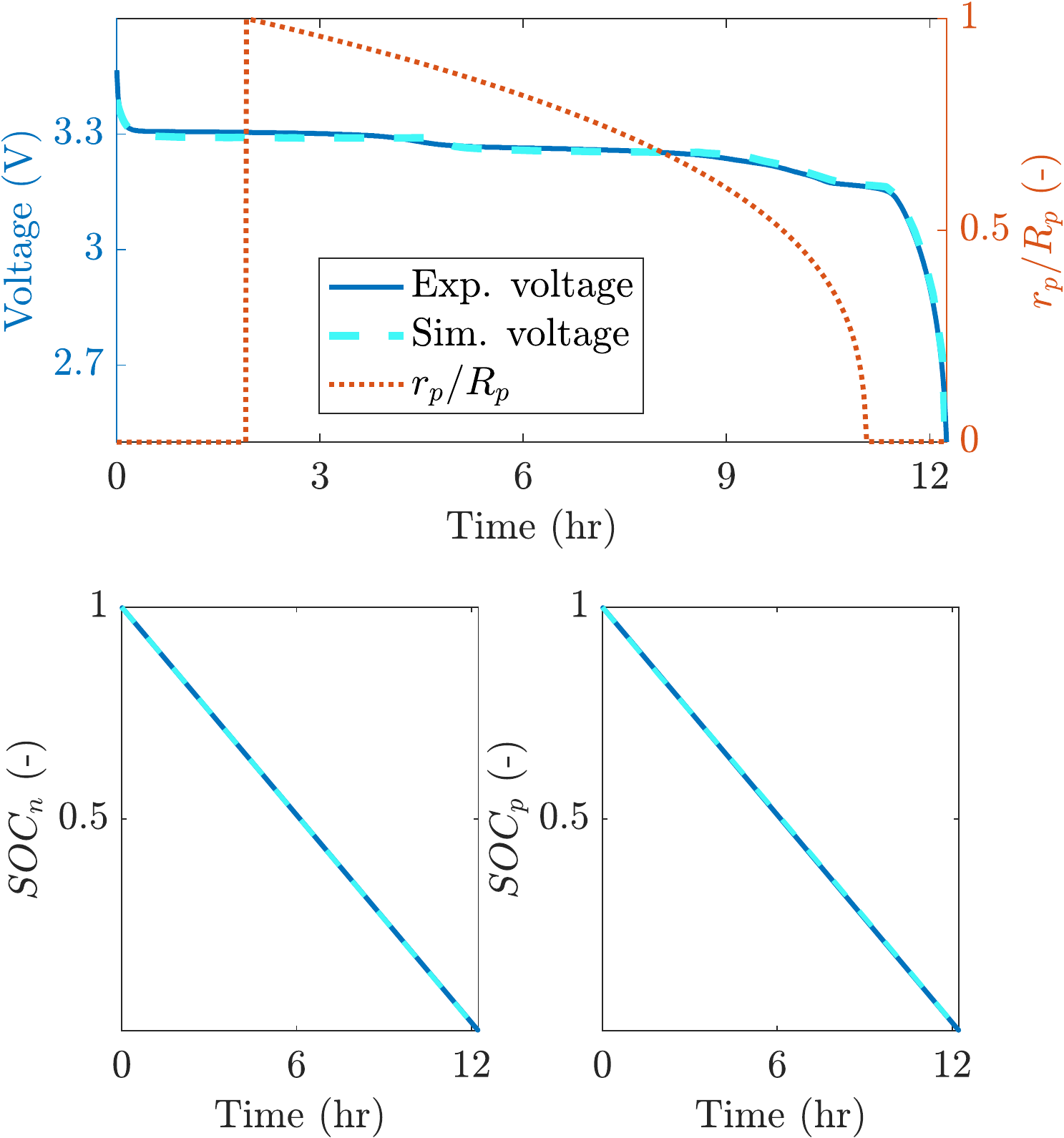}}\\
\subfloat[\textbf{Charge}]{\includegraphics[width=0.85\columnwidth]{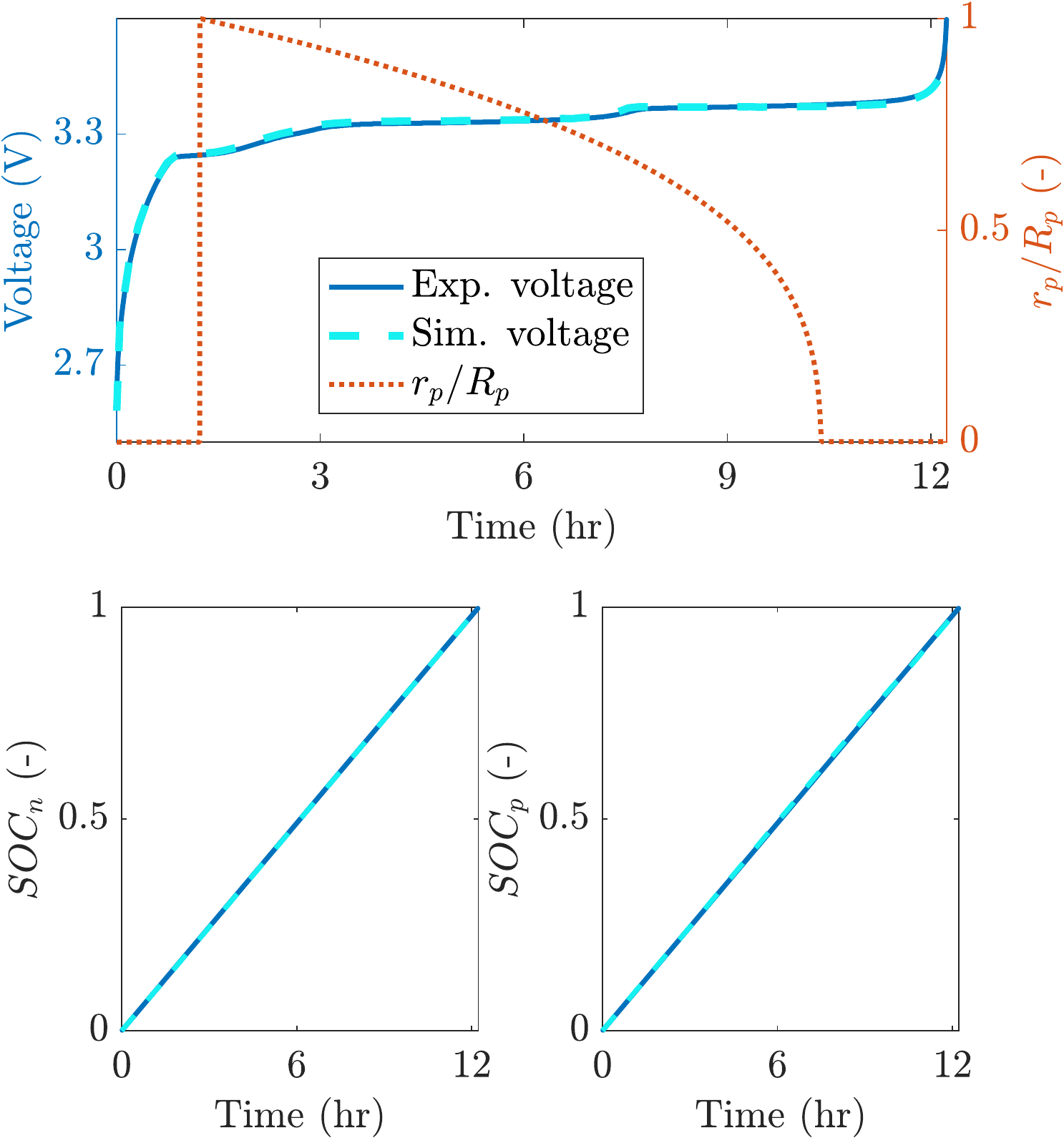}}
\caption{Comparison between C/12 experimental voltage and $SOC$ profiles with core-shell ESPM predicted voltage and $SOC$.  The behavior of the moving boundary $r_p/R_p$ is shown.}
\vspace{-1.25em}
\label{fig:di}
\end{figure}

Identification results are shown in Table \ref{identification_label}.  Bounds for the identified parameters (as well as values of the parameters that are not directly identified) are selected according to the available literature on LFP batteries \cite{li2015modeling}, \cite{prada2013simplified}, \cite{li2014current} and information provided by our industrial partner.  Bounds for $\theta_p^\alpha$ and $\theta_p^\beta$ ensure the transition from the one- to two-phase region, and vice versa, to occur at the beginning and at the end of the flat  OCP region.  According to Table \ref{identification_label},  only the identified $\theta_p^\beta$ hits the corresponding lower bound,  equal to 0.8.  Given that $\theta_p^\beta$ describes the transition from the one- to two-phase region,  further decreasing this bound is not physically meaningful since we would enter the two-phase region of the OCP.  
In Fig.~\ref{fig:di}, the  simulated voltage and $SOC$ profiles  are compared with C/12 experimental data for both charge and discharge conditions.  
The model perform well also with respect to the simulated open circuit voltage profile, and  as shown in Fig.  \ref{fig:di},  both the one- ($r_p/R_p = 0$) and two-phase ($r_p/R_p > 0$) regions are modeled accurately. As expected from Eq. \eqref{goveq:eq6},  the moving boundary reaches zero once the two-phase region is ended. 

\section{Conclusions}
In this paper,  a core-shell ESPM framework is  proposed.  The introduction of the core-shell dynamics in the modeling of the positive particle allows to describe the inherent intercalation and deintercalation process of lithium ions. 
The proposed core-shell ESPM is a first step for the development of reduced order models and electrode-based observers to be used in battery management system (BMS) applications.

\section*{Acknowledgment}
We thank LG Energy Solutions for their financial support. 

\bibliographystyle{IEEEtran}
\bibliography{biblio} 

\end{document}